\documentclass[aps,prd,twocolumn,showpacs,preprintnumbers,amsmath,amssymb]{revtex4}

\usepackage{graphicx}
\usepackage{dcolumn}
\usepackage{bm}
\usepackage{longtable}

\usepackage[all]{xy}
\usepackage{amssymb}
\usepackage{amsthm}
\usepackage{amsmath}
\usepackage{amsfonts}
\usepackage{fancybox}
\usepackage{amssymb}
\usepackage[dvips,dvipsnames,usenames]{color}

\newcommand{\beq}{\begin{eqnarray}}
\newcommand{\eeq}{\end{eqnarray}}

\begin{document}

\title{Quantum modified Regge-Teitelboim cosmology}

\author{Rub\'en Cordero}
\email{cordero@esfm.ipn.mx}
\affiliation{
Departamento de F\'\i sica, Escuela Superior de F\'\i sica y Matem\'aticas
del IPN, Unidad Adolfo L\'opez Mateos, Edificio 9, 07738, M\'exico, Distrito
Federal, M\'exico
}

\author{Miguel Cruz}
\email{miguel.cruz@ucv.cl}
\affiliation{Instituto de F\'\i sica, Pontificia Universidad Cat\'olica 
             de Valpara\'\i so, Casilla 4950, Valpara\'\i so, Chile \\
   and Departamento de F\'\i sica, Centro de Investigaci\'on y de Estudios Avanzados del IPN, Apdo Postal 14-740, 07000 M\'exico DF, M\'exico
          }

\author{Alberto Molgado}
\email{molgado@fc.uaslp.mx}
\affiliation{Facultad de Ciencias, Universidad Aut\'onoma de San Luis Potos\'{\i}, Av.
Salvador Nava S/N, Zona Universitaria CP 78290 San Luis Potos\'{\i}, SLP, M\'exico \\
and Dual CP Institute of High Energy Physics, M\'exico}

\author{Efra\'\i n Rojas}
 \email{efrojas@uv.mx}
\affiliation{Departamento de F\'\i sica, Facultad de F\'\i sica e Inteligencia
Artificial,    \\ Universidad Veracruzana, 91000 Xalapa, Veracruz, M\'exico}

\date{\today}
\begin{abstract}
The canonical quantization of the modified geodetic brane cosmology whi\-ch is
implemented from the Regge-Teitelboim model and the trace of the extrinsic 
curvature of the brane trajectory, $K$, is developed. As a second-order derivative 
model, on the grounds of the Ostrogradski Hamiltonian method and the Dirac's 
scheme for constrained systems, we find suitable first- and second-class 
constraints which allow for a proper quantization. We also find that the first-class 
constraints obey a sort of truncated Virasoro algebra. The effective quantum 
potential emerging in our approach is exhaustively studied where it shows 
that an embryonic epoch is still present. The quantum nucleation is sketched 
where we observe that it is driven by an effective cosmological constant.
\end{abstract}


\maketitle

%
\section{Introduction}
\label{sec1}
%

The modified geodetic brane gravity (MGBG) \cite{Cordero:2011zv} is an effective 
theory consisting of the Regge-Teitelboim model (RT), also named geodetic 
brane gravity (GBG)~\cite{Regge,pavsic,davidson0,davidson1,paston2} plus a geometric 
linear $K$ term which is, under certain conditions, responsible 
of mimic some features of the Dvali-Gabadadze-Porrati (DGP) theory~\cite{DGP, DGP1}. 
$K$ denotes the trace of the extrinsic curvature of the codimension one 
worldvolume swept out by a dynamical brane and, this is a measurement of how
the brane elements are oriented in the bulk. Apart from the cosmological constant,
the inclusion of this term into the RT model can be regarded as a minimum geometric 
extension that also leads to second-order equations of motion. In several 
frameworks such extrinsic curvature term has been studied: in the differential 
geometry of hypersurfaces~\cite{chen}, in the study of the bending and shape 
of phospholipid membranes~\cite{svetina} and, in the relativistic context, 
such a term has been considered to improve the extensible gravitational Dirac 
model of the electron~\cite{ot,electron,davidson3} as well as being considered an 
effective $4D$ field brane theory with possible applications in 
cosmology and particle physics~\cite{trodden1,trodden3}. 

The RT model was originally motivated  to describe our Universe in a point- or string-like 
fashion where our Universe is a $(3+1)$-dimensional extended object geodesically
floating in a fixed higher-dimensional bulk~\cite{Regge}. The associated brane-like 
cosmology was studied in~\cite{Davidson:1999fb}. Differential geometry aspects discussed
in~\cite{friedman:1965} show that to locally embed a metric on a surface, propagating 
in a $N$-dimensional flat background spacetime, the isometric 
embedding theorems dictate that $N=n(n+1)/2$ dimensions are required. 
In particular, for $n=4$, a ten-dimensional flat background is necessary. However, 
if the $(3+1)$-metric on the surface admits some Killing vector fields, $N$ can be
reduced significantly~\cite{rosen:1965}. The above arguments can also be applied 
when we include such $K$ term and, in this sense, it is attractive to implement 
MGBG for cosmology and in particular in quantum cosmology. Geometrically, MGBG is conformed 
only by the first three Lovelock brane invariants associated to the 
worldvolume~\cite{Cruz:2012bw}. In fact, the hypersurfaces described 
by such terms are characterized by a single degree of freedom associated only 
with the geometric configuration of the system~\cite{hambranes}. Relating to this fact there is a 
linkage with a peculiar set of second-order scalar field theories, free of 
ghosts, and considered as local modifications of gravity where the scalar 
degree of freedom $\pi$, the so-called {\it Galileon}, is a type of brane 
bending mode
\cite{Nicolis:2008in,deRham:2010eu,Goon:2011xf,Burrage:2011bt,Deffayet:2009mn,Deffayet:2009-2,Deffayet:2010,Fairlie:1992,trodden1,TRODDEN2}. 
There is thus a strong interest in all these classes of second-order Lagrangians, 
mainly for their potential applications at the cosmological level.

Within the minisuperspace framework it was shown in~\cite{Cordero:2011zv} that the 
introduction of the linear $K$ term provides an alternative mechanism to contrast the 
cosmological constant effects into the geodetic brane dynamics thus supplying 
a dynamical equivalence with the DGP model where the self-(non-self)-accelerated 
expansion of such  brane-like universe is mediated by the sign of the constant $\beta$ 
accompanying to the $K$ term. Conventionally this quantity is considered as the 
Gibbons-Hawking-York boundary term but, from the fact that we are not considering 
the bulk gravity to be dynamical, this second-order term is simply another possible 
geometrical invariant associated with the worldvolume which also leads to second-order 
equations of motion. A natural extension of the work developed in~\cite{Cordero:2011zv} 
is the one associated with the quantum approach in order to know some interesting 
features such as the brane nucleation of this type of universes. In this regards, 
the quantum theory associated with this brane model involves some technical troubles 
of considerable complexity where most part of the issues come from the linear dependence 
on the acceleration of the brane in the Lagrangian. Commonly this fact leads us to 
identify a divergence term that can be naively neglected without affecting the dynamics of 
the theory but, getting rid of such a term sometimes results harmful at Hamiltonian 
level as we cannot obtain constraints quadratic in the momenta in a straightforward 
way~\cite{Cordero:2011zv,Regge,davidson0,davidson1,Cordero:2009mj,Paul2013,Paul2013-2,davidson}.  
To obtain a form which would be appropriate for quantization, a robust prescription 
consists in maintain the second-order nature of the model and then to use a Hamiltonian 
development supported by a Dirac's procedure for second-order constrained 
systems~\cite{Dirac,Henneaux,Nesterenko1989}. 

This paper provides a companion to \cite{Cordero:2011zv} where the classical 
aspects of the MGBG within the minisuperspace framework are undertaken. After an 
Ostrogradski Hamiltonian treatment for constrained systems we find that, to obtain 
quadratic constraints in the momenta allowing for a canonical quantization, it 
is necessary to invoke a suitable canonical transformation followed of a gauge 
fixation. We thus obtain a Wheeler-DeWitt (WDW) type equation where an involved 
quantum potential emerges. An exhaustive analysis of this potential is 
done and it is found that a classically disconnected embryonic epoch (a characteristic 
feature of geodetic brane-like quantum cosmology) is still present. 
In fact, this embryo exists whenever the conserved energy $\Omega$, which is 
conjugate to the external time coordinate, is not zero. This quantum treatment paves 
the way to estimate the probability of creation for this brane-like universe. In this 
regards we observe that, for negative values of $\beta$ the creation of this type 
of accelerated universes is more probable, contrary to the case of positive values 
for $\beta$. Further, the nucleation rate for the particular case of a vanishing energy 
$\Omega$ is analyzed. It is shown that such probability resembles the one for general 
relativity by defining an effective cosmological constant in terms of the $\beta$ parameter.

The structure of the paper is as follows. In Sec.~\ref{sec2} we present a brief 
review of the modified geodetic brane gravity in order to set the physical stage. 
We specialize to a Friedmann-Robertson-Walker (FRW) metric on the brane embedded
in a flat background. This minimal embedding calls for only one extra dimension. 
Then we obtain an effective Lagrangian. In Secs.~\ref{sec3} and~\ref{sec4} we have 
succeeded in showing that by using an Ostrogradski Hamiltonian formulation besides 
a unique canonical transformation it is possible to obtain quadratic constraints in 
the physical momenta in order to pave the way to a naive canonical quantization. 
Further, we find a truncated Virasoro structure in the first-class constraint algebra. 
We establish a WDW equation in Sec.~\ref{sec5} where the emerging quantum potential 
is analyzed. In addition, the nucleation probability for this brane-like universe 
is calculated for a special case in Sec. \ref{sec6}. We finish in Sec.~\ref{sec7} 
with some conclusions of the work.

%
\section{Modified geodetic brane gravity}
\label{sec2}
%

Consider a three-dimensional dynamical brane. The $(3+1)$-dimensional worldvolume 
$m$, the brane-like universe, is embedded in a $(4+1)$-dimensional Minkowski background 
spacetime with metric $\eta_{\mu \nu}$ $(\mu , \nu=0,1,\ldots,4)$. We will 
assume that the dynamical variables are the embedding functions of $m$, $X^\mu 
(x^a)$, where $x^a$ are the worldvolume coordinates $(a,b=0,1,2,3)$. We 
construct the induced metric $ g_{ab} = \eta_{\mu\nu} e^\mu {}_a e^\nu{}_b 
:= e_a \cdot e_b$ and the extrinsic curvature $K_{ab} = - \eta_{\mu \nu} 
n^\mu \partial_a e^\nu {}_b$ where $e^\mu{}_a = \partial_a X^\mu$ are the tangent 
vectors to $m$ and $n^\mu$ is the normal vector defined uniquely (up to a sign) 
by $e_a \cdot n = 0$ and $n\cdot n=1$.

Under these geometric conditions, the MGBG theory for a three-dimensional brane is 
defined as~\cite{Cordero:2011zv}
\begin{equation}
S[X] = \int_m d^{4}x \sqrt{-g}\left(\frac{\alpha}{2}\mathcal{R} - \Lambda 
+ \beta K\right),
\label{eq:action}
\end{equation}
where ${\cal R}$ and $K=g^{ab}K_{ab}$ denote to the Ricci scalar and the mean 
extrinsic curvature of $m$, respectively. Here, $g:= \mbox{det} (g_{ab})$. 
In addition, $\alpha$ and $\beta$ are constants of dimensions $[L]^{-2}$ 
and $[L]^{-1}$ in Planck units, respectively, and $\Lambda$ is a positive 
cosmological constant defined on $m$. It is possible to consider some matter Lagrangians 
into the action~(\ref{eq:action}). Once matter is included, the form of the equations 
of motion is not affected~\cite{Cordero:2011zv,davidson1}. In this work we will 
only consider the cosmological constant effects, for simplicity.
The MGBG possesses as a main symmetry 
the invariance under reparametrizations of $m$. A variational procedure yields 
the equation of motion~\cite{Cordero:2011zv}
\begin{equation}
\alpha G_{ab} K^{ab} - \beta {\cal R} + \Lambda K =0,
\label{eq:eom}
\end{equation}
where $G_{ab}$ is the worldvolume Einstein tensor. This compact geometrically 
form represents a single second-order differential equation in derivatives of 
$X^\mu$ because of the presence of the extrinsic curvature tensor. This is so 
even though we have the presence of second-order derivative quantities in the 
action~(\ref{eq:action}) through the scalars ${\cal R}$ and $K$. Within a 
cosmological scenario the integration of the Eq.~(\ref{eq:eom}) gives rise to 
an important integration constant $\Omega$, which is nothing but the conserved 
bulk energy~\cite{Cordero:2011zv}.

For our purposes below, we embed a closed FRW universe in a Minkowski bulk 
$ds^{2}_{5} = -dt^{2} + da^{2} +a^{2}d\Omega^{2}_{3}$ where $d\Omega^{2}_{3} 
= d\chi^{2} + \sin^{2} \chi d\theta^{2} + \sin^{2}\chi \sin^{2}\theta d\phi^{2}$ 
is the unit three-sphere. By considering 
\begin{equation}
 X^\mu (x^a) = (t(\tau), a(\tau),\chi,\theta,\phi),
\label{eq:embedding}
\end{equation}
the induced metric is the FRW one
\begin{equation}
ds^{2}_{4} = -N^{2}d\tau^{2} + a^{2}d\Omega^{2}_{3},
\label{eq:induced}
\end{equation}
where $N^{2} = \dot{t}^{2} - \dot{a}^{2}$ and $a(\tau)$ being the scale factor. An 
overdot denotes differentiation with respect to $\tau$. Moreover, the unit 
normalized vector to $m$ is given by
\begin{equation}
n^\mu = \frac{1}{N} (\dot{a},\dot{t},0,0,0).
\label{eq:normal}
\end{equation}
This geometric configuration leads to
\begin{eqnarray}
\mathcal{R} &=& \frac{6\dot{t}}{N^{4}a^{2}}(a\ddot{a}\dot{t}-a\dot{a}\ddot{t} 
+ N^{2}\dot{t}),
\label{eq:R}
\\
K &=& \frac{1}{N^{3}}(\dot{t}\ddot{a} - \dot{a}\ddot{t}) + \frac{3\dot{t}}{aN}.
\label{eq:K}
\end{eqnarray}
From Eq. (\ref{eq:eom}) we have the equation of motion
\begin{equation}
\frac{d}{d \tau} \left( \frac{\dot{a}}{ \dot{t}} \right)
+ \frac{N^2\left( \dot{t}^2 - 3\bar{\Lambda} N^2 a^2 +
6 \bar{\beta} N a \dot{t} \right) }{a \dot{t}\left( 3\dot{t}^2 - \bar{\Lambda} N^2 a^2 +
6 \bar{\beta} N a \dot{t} \right)}= 0,
\end{equation}
where we have introduced the notation $\bar{\Lambda}^{2} := \Lambda/3\alpha$ and 
$\bar{\beta} := \beta/3\alpha$. In order to write down the action $S$ in analogy 
with analytical mechanics, we substitute first~(\ref{eq:R}) and~(\ref{eq:K}) 
into~(\ref{eq:action}), then after an integration over the spatial coordinates 
the action reduces to $S = 6\pi^{2} \int d\tau L$ where the Lagrangian function reads
\begin{equation}
L = \frac{a\dot{t}}{N^{3}}(a\ddot{a}\dot{t} - a\dot{a}\ddot{t} + N^{2}\dot{t}) 
- Na^{3}\bar{\Lambda}^{2} + \frac{a^{3}\bar{\beta}}{N^{2}}(\dot{t}\ddot{a} - 
\dot{a}\ddot{t}) + 3a^{2}\bar{\beta}\dot{t}.
\label{eq:lagrangian}
\end{equation}
Notice a linear dependence in the accelerations of the coordinates $a(\tau)$ and 
$t(\tau)$. In the fashion (\ref{eq:lagrangian}) we infer that the configuration 
space is spanned by $\left\lbrace t,a,\dot{t},\dot{a} \right\rbrace$.

Certainly, $L$ can be split as $L = L_{b} + L_{d}$ where
\begin{equation}
L_{b} = \frac{d}{d\tau}\left[\frac{a^{2}\dot{a}}{N} + a^{3}\bar{\beta}\mbox{arctanh}
\left(\frac{\dot{a}}{\dot{t}}\right)\right],
\label{bound}
\end{equation}
and
\begin{equation}
L_{d}= -\frac{a\dot{a}^{2}}{N} + aN(1-a^{2}\bar{\Lambda}^{2}) + 3a^{2}
\bar{\beta}\left[\dot{t} -\dot{a} \mbox{arctanh}\left(\frac{\dot{a}}{\dot{t}} 
\right)\right].
\label{dynamical}
\end{equation}
$L_{b}$ denotes a boundary Lagrangian term which produces no dynamics so that we 
can neglect it without affecting the equations of motion. To Hamiltonian purposes 
this strategy sometimes is not suitable if we want to obtain quadratic constraints 
in the momenta unless we introduce auxiliary field variables which extends the canonical 
analysis~\cite{Cordero:2011zv,davidson0,davidson1,Paul2013-2,paston1}. Fortunately, from 
a second-order derivative viewpoint, a robust prescription lies in to maintain 
intact the Lagrangian~(\ref{eq:lagrangian}) followed by an Ostrogradski Hamiltonian 
approach as we will see shortly.

%
\section{Ostrogradski Hamiltonian approach}
\label{sec3}
%

Given the second-order Lagrangian~(\ref{eq:lagrangian}), we must note that 
due to its linear dependence on the acceleration, this is degenerate but 
stable, as we will discuss below by using the Dirac's framework needed to 
deal with constrained systems~\cite{Dirac,Henneaux,Nesterenko1989}. First, 
by following the Ostrogradski construction~\cite{Ostro:1850} we identify 
that the phase space is spanned by $\left\lbrace t,p_t,a,p_a; \dot{t},P_t, 
\dot{a}, P_a \right\rbrace$ where the conjugate momenta to the velocities 
$\lbrace{\dot{t}, \dot{a}\rbrace}$ are given by
\begin{subequations}
\begin{eqnarray}
P_{t} &=& \frac{\partial L}{\partial \ddot{t}} 
= -\frac{a^{2}\dot{a}}{N^{3}}(\dot{t} + Na\bar{\beta}),
\label{Pt}
\\
P_{a} &=& \frac{\partial L}{\partial \ddot{a}} 
= \frac{a^{2}\dot{t}}{N^{3}}(\dot{t} + Na\bar{\beta}),
\label{Pa}
\end{eqnarray}
\end{subequations}
and
\begin{subequations}
\begin{eqnarray}
p_{t} &=& \frac{\partial L}{\partial \dot{t}} - \frac{d}{d\tau}\left(\frac{\partial L}{\partial 
 \ddot{t}}\right)  
\nonumber \\ 
&=& \frac{a\dot{t}}{N^{3}}[\dot{a}^{2} + N^{2}(1-a^{2}\bar{\Lambda}^{2}) 
 + 3\bar{\beta}Na\dot{t}] =: -\Omega, 
\label{energy}
\\
p_{a} &=& \frac{\partial L}{\partial \dot{a}} - \frac{d}{d\tau}\left(\frac{\partial L}{\partial 
\ddot{a}}\right)  
\nonumber 
\\ 
&=& -\frac{a\dot{a}}{N^{3}}[\dot{a}^{2} + N^{2}(1-a^{2}\bar{\Lambda}^{2}) 
+ 3\bar{\beta}Na\dot{t}],
\label{pa}
\end{eqnarray}
\end{subequations}
being the conjugate momenta to the position variables $\lbrace{t, a\rbrace}$.
It is worthwhile to mention that $p_{t}$ is not affected by the surface Lagrangian 
term~(\ref{bound}) because it is nothing but the conserved bulk energy 
$\Omega$~\cite{Cordero:2011zv} which parametrizes the deviation from the Einstein 
limit whenever $\beta \to 0$. With regards the momentum $p_{a}$, it is composed by 
two contributions, $p_a = \mathbf{p}_{a} + \mathfrak{p}_{a}$. The momentum
$\mathbf{p}_a$ is associated to the equivalent dynamical theory defined by 
(\ref{dynamical}) whereas $\mathfrak{p}_a$ is related to the boundary Lagrangian 
term (\ref{bound})~\cite{electron,Cordero:2009mj}. Explicitly, they are 
given by
\begin{eqnarray}
\mathbf{p}_{a} &=& -\frac{a\dot{a}}{N^{3}}\left[\dot{a}^{2} 
+ N^{2}(3 - a^{2}\bar{\Lambda}^{2})\right] 
\nonumber
\\
&-& 3a^{2}\bar{\beta}\left[\frac{\dot{a}\dot{t}}{N^{2}}
+ \mbox{arctanh}\left(\frac{\dot{a}}{\dot{t}}\right)
\right],
\label{eq:pda}
\\
\mathfrak{p}_{a} &=& \frac{2a\dot{a}}{N} 
+ 3a^{2}\bar{\beta}\mbox{arctanh}\left(\frac{\dot{a}}{\dot{t}}
\right).
\label{eq:pba}
\end{eqnarray}
In this sense, $p_t = \mathbf{p}_t$. For our analysis below, it is crucial to maintain $p_a$ in terms 
of the two pieces, (\ref{eq:pda}) and (\ref{eq:pba}).

The canonical Hamiltonian which defines the appropriate phase space is provided by the Ostrogradski
construction~\cite{Nesterenko1989,Ostro:1850}
\begin{eqnarray}
H_0 &=& P \cdot \ddot{X} +  p \cdot \dot{X}  - L,
\nonumber
\\
&=&  p\cdot \dot{X} + N\left(a^{3}\bar{\Lambda}^{2}-\frac{1}{a^{3}}N^{2}P^{2} + \bar{\beta}^{2}a^{3} 
\right.
\nonumber
\\
&-& \left. \bar{\beta}a^{2}\frac{\dot{t}}{N} \right).
\end{eqnarray}
The definition of the momenta (\ref{Pt}) and (\ref{Pa}) gives rise to two 
primary linear constraints in the momenta
\begin{eqnarray}
\phi_1 &=& P_t +  \frac{a^{2}\dot{a}}{N^{3}}(\dot{t} + Na\bar{\beta}) \approx 0,
\label{eq:prim-1}
\\
\phi_2 &=& P_a -  \frac{a^{2}\dot{t}}{N^{3}}(\dot{t} + Na\bar{\beta}) \approx 0,
\label{eq:prim-2}
\end{eqnarray}
which can be collected in the compact form $\phi_\mu = P_\mu - \frac{a^2 (\dot{t} 
+ \bar{\beta} N a)}{N^2}n_\mu$. Here, $\approx$ stands for weak equality in the 
Dirac's scheme for constrained systems. By projecting  $\phi_\mu$ along the velocity
vector as well as the unit normal vector to $m$ at a fixed time, we can
obtain a more suitable set of primary constraints\footnote{This fact is supported by using
an existing completeness relation in the geometry of deformations for branes, namely
$\eta^{\mu \nu} = n^\mu n^\nu - \eta^\mu \eta^\nu + h^{AB} \epsilon^\mu{}_A \epsilon^\nu{}_B$. 
See Ref.~\cite{Cordero:2009mj}.}
\begin{subequations}
\begin{eqnarray}
\varphi_1 &=& P_t \dot{t} + P_a \dot{a} = P \cdot \dot{X} \approx 0,
\label{eq:Prim-1}
\\
\varphi_2 &=& N (P \cdot n) - \frac{a^2}{N}(\dot{t} + \bar{\beta} N a) \approx 0,
\label{eq:Prim-2}
\end{eqnarray}
\end{subequations}
so that the total Hamiltonian is $H_T = H_0 + u^1 \varphi_1 + u^2 \varphi_2$ where 
$u^{1,2}$ are Lagrange multipliers enforcing (\ref{eq:Prim-1}) and (\ref{eq:Prim-2}). 
Apparently, in $H_0$ the linear dependence in the momentum $p_\mu$ leads to the so-called 
Ostrogradski linear instability~\cite{tolley} which force to the manifestation of ghost 
degrees of freedom but this appearance is however deceptive as we will show later on.

By using the extended Poisson bracket (PB) between two phase space functions, $f$ and $g$,
\begin{eqnarray}
\left\lbrace f,g  \right\rbrace &=& \frac{\partial f}{\partial t} \frac{\partial g}{\partial p_t}
+ \frac{\partial f}{\partial a} \frac{\partial g}{\partial p_a} 
+ \frac{\partial f}{\partial \dot{t}} \frac{\partial g}{\partial P_t}
+ \frac{\partial f}{\partial \dot{a}} \frac{\partial g}{\partial P_a}
\nonumber
\\
&-& (f \leftrightarrow g),
\label{eq:PB}
\end{eqnarray}
as befits a second-order derivative theory, we obtain that secondary constraints are generated
by the consistency relations $\dot{\varphi}_{1,2} = \left\lbrace \varphi_{1,2},H_T  \right\rbrace
\approx 0$. Thus, we obtain two secondary constraints
\begin{subequations}
\begin{eqnarray}
\varphi_3 &=& H_0 \approx 0,
\label{eq:second-1}
\\
\varphi_4 &=& p_t \dot{a} + p_a \dot{t} = N (p \cdot n) \approx 0.
\label{eq:second-2}
\end{eqnarray}
\end{subequations}
There are not tertiary constraints. The relevant physical information is obtained when primary and
secondary constraints are separated into first- and second-class constraints ${\cal F}$'s
and ${\cal S}$'s, respectively. For our case we have
\begin{subequations}
\begin{eqnarray}
\mathcal{F}_{1} &=& P\cdot \dot{X} \approx 0,
\label{f1}
\\
\mathcal{F}_{2} &=& \left( \frac{N \Theta}{a^2 \Phi}\right) \varphi_2 + H_{0} 
\approx 0,
\label{f2}
\\
\mathcal{S}_{1} &=& 
\varphi_2 \approx 0,
\label{s1}
\\
\mathcal{S}_{2} &=&  \varphi_4
\approx 0,
\label{s2}
\end{eqnarray}
\end{subequations}
where $\Theta:= N^2 \Omega + 2N \bar{\Lambda}^2 a^3 \dot{t} - 3 \bar{\beta} a^2 \dot{t}^2$ 
and  $\Phi :=3\dot{t}^2 - N^2 a^2 \bar{\Lambda}^2 + 6 \bar{\beta} N a \dot{t} $. Note that
by imposing the condition~(\ref{s1}), the total Hamiltonian $H_T$ is replaced then by the 
first-class Hamiltonian
\begin{equation}
H = {\cal F}_2 + u^1\,{\cal F}_1.
\label{eq:H}
\end{equation}
In fact, the evolution predicted by $H_T$ and $H$ is the same~\cite{Henneaux}. To complement 
our canonical approach we must replace the PB with the Dirac bracket (DB) defined by
\begin{equation}
\left\lbrace f,g \right\rbrace^{*} := \left\lbrace f,g \right\rbrace - \left\lbrace f,{\cal S}_i 
\right\rbrace {\cal S}^{-1} _{ij} \left\lbrace {\cal S}_j,g \right\rbrace,
\label{eq:DB}
\end{equation}
where ${\cal S}_{ij} ^{-1}$ denotes the inverse elements of the second-class constraint matrix 
${\cal S}_{ij} := \left\lbrace {\cal S}_i, {\cal S}_j \right\rbrace$, $(i,j = 1,2.)$. Explicitly
\begin{equation}
({\cal S}_{ij}) = - \frac{a \Phi}{N} \left(
\begin{array}{cc}
0 & 1
\\
-1 & 0
\end{array}
\right).
\end{equation}
In view of the Dirac's constraint method, we must consider the second-class constraints
to vanish strongly which helps to eliminate the part proportional to $\varphi_2$ in (\ref{f2})
leading thus to a simplified expression for ${\cal F}_2$. The counting of the physical degrees 
of freedom (dof) is straightforward~\cite{Henneaux}: $\mbox{dof} =[8 - 2\times 2 - 2]/2 = 1$. This 
geometrical dof is the one that account for the brane bending mode in our approach 
and related to $a(\tau)$. 

As for the first-class constraint algebra, the DB between the ${\cal F}_1$ 
and the reduced ${\cal F}_2$ reads
\begin{equation}
\left\lbrace {\cal F}_i , {\cal F}_j \right\rbrace^{*} = -\epsilon_{ij} {\cal F}_2 ,
\quad \qquad i,j = 1,2,
\label{eq:algebra}
\end{equation}
with $\epsilon_{ij}$ being the Levi-Civita symbol such that $\epsilon_{12}=1$. Now, this 
expression suggests to introduce the notation $L_0 := {\cal F}_1$ and $L_1 := {\cal F}_2$. The 
relation (\ref{eq:algebra}) transforms then into
\begin{equation}
 \left\lbrace L_m , L_n \right\rbrace^{*} = (m-n) L_{m+n}, \quad m=0,n=1;
\label{virasoro}
\end{equation}
which characterizes to a truncated Virasoro algebra~\cite{Paul2013,Paul2013-2,Ho2003}. 
We claim, based on some models recently studied~\cite{Cordero:2011zv,Cordero:2009mj,electron}, 
that this is a symmetry inherited by all Lovelock brane Lagrangians 
characterized by a linear dependence in the acceleration of the brane. It will be reported
elsewhere. In summary, we have two first-class constraints~(\ref{f1}) and (\ref{f2}) 
reflecting the invariance under reparametrizations of the worldvolume $m$ and obeying 
a truncated Virasoro algebra, (\ref{virasoro}). On the other hand, we have two second-class 
constraints~(\ref{eq:second-1}) and (\ref{eq:second-2}) that signal the fact that 
the velocities and their conjugate momenta are not physical fields.

\section{Canonical transformation and gauge fixing}
\label{sec4}

In order to get quadratic constraints in the momenta we need to re-express the 
set of constraints~(\ref{f1}-\ref{s2}) in a convenient way.  To do this, we consider 
the following canonical transformation (CT)~\cite{electron,Cordero:2009mj}
\begin{eqnarray}
N&:=& \sqrt{\dot{t}^{2}-\dot{a}^{2}},
\label{eq:N}
\\
\Pi_{N} &:=& \frac{1}{N}(P\cdot \dot{X}), 
\label{canonical1}
\\
v &:=& -\left[N(P\cdot n) - \frac{a^{2}}{N} (\dot{t} + \bar{\beta} N a ) \right], 
\label{eq:nu}
\\
\Pi_{v} &:=& \mbox{arctanh}\left(\frac{\dot{a}}{\dot{t}}\right),
\label{canonical2}
\end{eqnarray}
together with the transformation $X^{\mu} = X^{\mu}$ and ${\bf p}_\mu = p_\mu - 
\mathfrak{p}_\mu$. This canonical transformation preserves the Poisson 
bracket structure in the sense that
\begin{equation}
\lbrace N, \Pi_{N} \rbrace = 1 = \lbrace v , \Pi_v \rbrace \quad
\mbox{and}\quad \left\lbrace X^\mu , \mathbf{p}_\nu \right\rbrace = \delta^\mu {}_\nu.  
\end{equation}
In addition, this CT dictates that the velocity vector can be written as
\begin{equation}
\dot{X}^\mu = N (\cosh \Pi_v, \sinh \Pi_v,0,0,0), 
\label{eq:velocity}
\end{equation}
while the momenta (\ref{energy}) and (\ref{pa}) become
\begin{widetext}
\begin{eqnarray}
\mathbf{p}_{t} &=& a[\sinh^{2} \Pi_v + (1-a^{2}\bar{\Lambda}^{2}) + 3\bar{\beta}a\cosh \Pi_v]\cosh 
\Pi_v = -\Omega, 
\label{newenergy}
\\
p_{a} &=&  -a[\sinh^{2} \Pi_{v} + (1-a^{2}\bar{\Lambda}^{2}) + 3\bar{\beta}a\cosh \Pi_v]\sinh 
\Pi_v = \Omega \tanh \Pi_v,
\label{newpa}
\end{eqnarray}
\end{widetext}
or, in a more compact form
\begin{equation}
p_\mu = \Omega (-1, \tanh \Pi_v,0,0,0). 
\end{equation}
With regards to the momenta (\ref{eq:pda}) and (\ref{eq:pba}) we have
\begin{eqnarray}
\mathbf{p}_a &=& \left\lbrace -a \left[\sinh^{2} \Pi_v + (3 - a^{2}\bar{\Lambda}^{2}) \right] 
- 3\bar{\beta}a^{2}\cosh \Pi_v \right\rbrace \times
\nonumber
\\
&& \quad \sinh \Pi_v - 3\bar{\beta} a^2 \Pi_v, 
\label{newphasepa}
\\
\mathfrak{p}_a &=& 2a\sinh \Pi_v + 3\bar{\beta} a^2 \Pi_v.
\label{newphasepba}
\end{eqnarray}
It remains to express the momenta (\ref{Pt}) and (\ref{Pa}) in terms of the new phase space 
variables for completeness
\begin{subequations}
\begin{eqnarray}
P_t &= \Pi_N \cosh \Pi_v + \frac{1}{N} (v - \bar{\beta} a^3 - a^2 \cosh \Pi_v) \sinh \Pi_v,
\nonumber
\\
P_a &= - \Pi_N \sinh \Pi_v - \frac{1}{N} (v - \bar{\beta} a^3 - a^2 \cosh \Pi_v) \cosh \Pi_v.
\nonumber
\end{eqnarray}
\end{subequations}
Thus, we are able now to rewrite the first-class constraints as follows,
\begin{widetext}
\begin{subequations}
\begin{eqnarray}
\mathcal{F}_{1} &=& N\Pi_{N},
\label{eq:N-constraint}
\\
\mathcal{F}_{2} &=& N\left[ \mathbf{p}_{t}\cosh \Pi_v + (\mathbf{p}_{a} + \mathfrak{p}_{a})\sinh \Pi_v 
+ a^{3}\bar{\Lambda}^{2} + \frac{1}{a^{3}}N^{2}\Pi^{2}_{N} - a\cosh^{2}\Pi_v 
- 3\bar{\beta}a^{2}\cosh \Pi_v
\right. 
\nonumber
\\
&-& \left.  \frac{1}{a^{3}}v (v -2a^{2}\cosh \Pi_v - 2\bar{\beta}a^{3})\right],
\label{constraint}
\end{eqnarray}
\end{subequations}
\end{widetext}
where we have used the relation $N^{2}P^{2} = -(P\cdot \dot{X})^{2} + N^{2}(P\cdot 
n)^{2}$~\cite{Cordero:2009mj} together with (\ref{eq:N}-\ref{canonical2}). Similarly, regarding  
the second-class constraints we have
\begin{eqnarray}
\mathcal{S}_{1} &=& v , 
\label{second1}
\\
\mathcal{S}_{2} &=& N(\mathfrak{p}_{a} - 2a\sinh \Pi_v - 3\bar{\beta}a^{2} \Pi_v)\cosh 
\Pi_v .
\label{second2}
\end{eqnarray}
We observe immediately that ${\cal S}_2$ reduces to the definition of the momenta provided by
the Lagrangian (\ref{bound}).

The fact that we have two first-class constraints signals that we have the freedom
to choose two gauge conditions. We impose the so-called {\it cosmic gauge}
\begin{equation}
C_1 = N - 1 \approx 0, 
\label{gauge1}
\end{equation}
and
\begin{equation}
C_2 = \cosh \Pi_v - \sqrt{\gamma} a \bar{\Lambda} \approx 0,
\label{gauge2}
\end{equation}
where $\gamma = \gamma(a)$. From the expression $C_{2}$ and the definition of the momenta 
$p_{t}$, Eq. (\ref{energy}), we see that $\gamma$ must obey the rather involved equation 
\begin{equation}
\gamma \left(\gamma - 1 + 3\sqrt{\gamma}\frac{\bar{\beta}}{\bar{\Lambda}}\right)^{2} = 
\frac{\Omega^ {2}}{a^{8}\bar{\Lambda}^{6}}.
\label{gamma}
\end{equation} 
Inclusion of the function $\gamma(a)$ will be helpful in order to introduce the conserved 
energy $\Omega$ within our quantum approach. This gauge condition is totally equivalent to 
the expression $\sqrt{\dot{a}^2 + N^2} - \sqrt{\gamma} N a\bar{\Lambda} = 0$ where we have 
used the time component of (\ref{eq:velocity}) and the new canonical variable $N$ given 
by~(\ref{eq:N}).
The relations (\ref{gauge1}) and (\ref{gauge2}) completely fix the gauge freedom
associated to the invariance under reparametrizations. These gauge conditions are 
good enough since the square matrix $\left( \left\lbrace C_{1,2}, {\cal F}_{1,2} 
\right\rbrace \right) $ results nondegenerate in the constraint surface. Indeed, taking
advantage that the symplectic structure as defined in Eq. (19) holds 
 when evaluated with respect to the new canonical variables (\ref{eq:N}-\ref{canonical2}) 
 together with $X^\mu$ and $\mathbf{p}_\mu$, we have
\begin{eqnarray}
\left\lbrace  C_1, {\cal F}_1 \right\rbrace &=& C_1 + 1, \ \ \left\lbrace  
C_1, {\cal F}_2 \right\rbrace = 0,
\\
\left\lbrace  C_2, {\cal F}_1 \right\rbrace &=& 0, \ \ \ \ \ \ \ \ \  \, \left\lbrace  C_2, 
{\cal F}_2 \right\rbrace = G (a,N,v,\Pi_v),
\end{eqnarray}
where $G$ is a nonvanishing function\footnote{This function is given by
the relation $G = \left[ \frac{2}{a^3} \left( v - a^2 \cosh \Pi_v 
- \bar{\beta} a^3 \right) - \bar{\Lambda} \left( \sqrt{\gamma} + 
\frac{a}{2\sqrt{\gamma}} \frac{\partial \gamma}{\partial a}
\right) \right] 
N \sinh \Pi_v.$}.  Hence, the condition $\mathrm{det}\left( 
\left\lbrace C_i, {\cal F}_j \right\rbrace \right) \neq 0 $
with $i,j = 1,2,$ is  fulfilled.

The key point now is to express the physical momenta, $\mathbf{p}_t$ and $\mathbf{p}_a$, in terms 
of the gauge fixing conditions. From Eqs.~(\ref{newenergy}),~(\ref{newphasepa}) we have
\begin{widetext}
\begin{eqnarray}
\mathbf{p}_t &=&  a[\sinh^{2} \Pi_v + (1-a^{2}\bar{\Lambda}^{2}) + 3\bar{\beta}a\cosh \Pi_v]\cosh 
\Pi_v, 
\\
- \left( \mathbf{p}_a + 3\bar{\beta} a^2 \Pi_v \right) &=& a \left( \cosh^2 \Pi_v
- a^2 \bar{\Lambda}^2 + 2 + 3 \bar{\beta} a \cosh \Pi_v \right) \sinh \Pi_v.
\end{eqnarray}
\end{widetext}
Now, by considering the gauge condition~(\ref{gauge2}) we have
\begin{eqnarray}
\cosh \Pi_v &=& \frac{\mathbf{p}_t}{a\left[(\gamma - 1)a^{2}\bar{\Lambda}^{2} + 3\bar{\beta}\sqrt{\gamma}
a^{2}\bar{\Lambda}\right]},\\
\sinh \Pi_v &=& \frac{\left(\mathbf{p}_{a} + 3\bar{\beta}a^{2}\Pi_v\right)}{a\left[(\gamma - 1)a^{2}
\bar{\Lambda}^{2} + 2 + 3\bar{\beta}\sqrt{\gamma}a^{2}\bar{\Lambda}\right]}.
\end{eqnarray}
When we insert these expressions in the constraints~(\ref{eq:N-constraint}) and~(\ref{constraint}), 
these become 
\begin{widetext}
\begin{eqnarray}
\chi_1 &=& N\Pi_{N},
\label{Nconstraint}
\\
\chi_2 &=& -\frac{N}{a\left[(\gamma - 1)a^{2}\bar{\Lambda}^{2} + 2 
+ 3\bar{\beta} \sqrt{\gamma}a^{2}\bar{\Lambda}\right]} \left\lbrace 
\left(\mathbf{p}_{a} + 3\bar{\beta}a^{2}\Pi_{\nu}
\right)^{2} \right.
\nonumber
\\
&-& \left. a \left(\frac{p_{t}^{2}}{a\left[(\gamma - 1)a^{2}\bar{\Lambda}^{2} 
+ 3\bar{\beta}\sqrt{\gamma}a^{2} \bar{\Lambda}\right]}  
+ \frac{1}{a^{3}}N^{2}\Pi_{N}^{2} +  2a(\gamma a^{2}\bar{\Lambda}^{2} - 1) 
- (\gamma - 1)a^{3}\bar{\Lambda}^{2} - 3\bar{\beta}a^{3}\sqrt{\gamma}
\bar{\Lambda}\right) \times
\right.
\nonumber
\\
&& \left. \quad  \left[(\gamma - 1)a^{2}\bar{\Lambda}^{2} + 2 
+ 3\bar{\beta}\sqrt{\gamma}a^{2}\bar{\Lambda}\right] \right\rbrace,
\label{constraintquadratic}
\end{eqnarray}
\end{widetext}
where in $\chi_2$ is reflected a quadratic dependence in the momenta of the theory. 
Thus, following the Dirac's formalism for constrained systems, once we fix the 
gauge freedom we are left with a pure second-class system $\left( \chi_1,\chi_2, 
\chi_3 := {\cal S}_1, \chi_4 := {\cal S}_2 \right)$. 
These second-class constraints are regarded as simple identities expressing some 
dynamical variables in terms of others and all the equations of the theory are formulated 
in terms of the DB. We have learned then that a canonical transformation resolves the 
conflict of obtaining an appropriate form for quantization as remarked in the Introduction. 
As a byproduct, note that we have removed the Ostrogradski linear instability by removing 
structures associated to higher order terms. 

With regards to the DB definition~(\ref{eq:DB}) we have that
\begin{equation}
\left\lbrace N,\Pi_N \right\rbrace^* = \left\lbrace t ,\mathbf{p}_t \right\rbrace^* 
= \left\lbrace a ,\mathbf{p}_a \right\rbrace^* = 1  \,\, \mbox{and} \,\, \left\lbrace v, \Pi_v 
\right\rbrace^* = 0.
\label{eq:basicDB}
\end{equation}
The fact that $\left\lbrace v, \Pi_v \right\rbrace^* = 0$ tell us that the pair $\left( v, \Pi_v 
\right)$ does not describe a true physical degree of freedom since the resulting algebra associated
with the $\left( N, \Pi_N, t, \mathbf{p}_t , a, \mathbf{p}_a \right)$ sector of the theory is the one
that is closed under the DB~\cite{Henneaux}.

%
\section{Modified geodetic brane quantum cosmology}
\label{sec5}
%

The transition to the quantum mechanical scheme is carried out in the standard way. 
The structure of the DB is replaced with that of a commutator. Therefore, the correspondence
rule $i\widehat{\left\lbrace A, B\right\rbrace^{*}} = [\hat{A},\hat{B}]$ for two quantum 
operators $\hat{A}$ and $\hat{B}$ (modulo factor ordering and, $\hbar=1$) with $v$ and 
$\Pi_v$ replaced by the zero operator, yield a satisfactory theory in which only the 
canonical pairs $\left( N, \Pi_N\right)$, $\left( t, \mathbf{p}_t \right)$ and $\left(
a, \mathbf{p}_a \right)$ are realized as nontrivial quantum operators. Hence, we are now 
equipped to canonically quantize our model and according to the usual procedure we claim 
first that in a coordinate representation 
\begin{eqnarray}
\mathbf{p}_t &\longrightarrow& \widehat{\mathbf{p}}_t = -i\frac{\partial}{\partial t},
\label{pt-operator}
\\
\mathbf{p}_a &\longrightarrow& \widehat{\mathbf{p}}_{a} = -i\frac{\partial}{\partial a},
\label{pa-operator}
\\
\Pi_N &\longrightarrow& \widehat{\Pi}_N = - i \frac{\partial}{\partial N}.
\label{PiN-operator}
\end{eqnarray}
With this prescription we can consistently enforce our constraints as operator equations. 
The Hamiltonian~(\ref{eq:H}), composed now by the second-class constraints $\chi_1$ and 
$\chi_2$, is the one which is to be quantized. Thus, the physical states, $\Psi$, for our 
constrained system are those anihilated by the operator equations 
\begin{eqnarray}
\widehat{\chi}_1 \Psi &=& 0,
\label{eq:DN}
\\
\widehat{\chi}_2 \Psi &=& 0. 
\label{eq:schrodi}
\end{eqnarray}
Here, for simplicity we will choose a trivial factor ordering which allow us to get 
rid of the denominator in~(\ref{constraintquadratic}) (see the discussion, for example, 
in~\cite{electron}).

Thus, by inserting (\ref{pt-operator}-\ref{PiN-operator}) into (\ref{Nconstraint}) and
(\ref{constraintquadratic}), acting on $\Psi$, we obtain the differential equations
\begin{widetext}
\begin{eqnarray}
\widehat{\chi}_1\Psi &=&  -iN\frac{\partial \Psi}{\partial N} = 0,
\label{eq:op-N} 
\\
\widehat{\chi}_2 \Psi &=& -\frac{N}{a\left[(\gamma - 1)a^{2}\bar{\Lambda}^{2} + 2 
+ 3\bar{\beta} \sqrt{\gamma}a^{2}\bar{\Lambda}\right]} \left\lbrace -
\left( \frac{\partial^2}{\partial a^2} 
\right) \right.
\nonumber
 \\
& -& \left. a \left[ \frac{1}{a\left[(\gamma - 1)a^{2} \bar{\Lambda}^{2} + 3\bar{\beta} 
\sqrt{\gamma}a^{2}\bar{\Lambda}\right]}\left(\frac{\partial^{2}}{\partial t^{2}}\right)   
- \frac{1}{a^{3}}N^{2} \left( \frac{\partial^{2}}{\partial N^{2}} \right) 
+ 2a(\gamma a^{2}\bar{\Lambda}^{2} - 1) 
- (\gamma - 1)a^{3}\bar{\Lambda}^{2} - 3\bar{\beta}a^{3}\sqrt{\gamma}\bar{\Lambda}
\right] \times
\right. 
\nonumber
\\
& &\left. \left[
(\gamma - 1)a^{2}\bar{\Lambda}^{2} + 2 + 3\bar{\beta}\sqrt{\gamma}a^{2}
\bar{\Lambda}\right] \right\rbrace \Psi = 0.
\label{eq:maciza}
\end{eqnarray}
\end{widetext}
Eq. (\ref{eq:op-N}) entails that the physical states $\Psi$ have not a $N$ 
dependence. Consequently, we are left with equation (\ref{eq:maciza}) which results to be the 
Schr\"odinger-like equation that we are looking for, as it was expected.

We assume then that $\Psi$ is represented in the usual manner as  $\Psi(a,t):= \psi(a)
e^{-i \Omega t}$ in agreement with the classical definition of $\Omega$. Substituting $\Psi$ in 
(\ref{eq:maciza}) followed of a lengthy but straightforward computation, we find after removal 
of the exponential term that $\psi(a)$ satisfies the WDW type equation 
\begin{equation}
\left[-\frac{\partial^{2}}{\partial a^{2}} + U(a)\right]\psi(a) = 0,
\end{equation}
which looks like a zero-energy Schr\"{o}dinger equation with the quantum potential 
\begin{eqnarray}
U(a) &=& a^{2}\left[(\gamma -1)a^{2}\bar{\Lambda}^{2} + 2 + 3\bar{\beta}
\sqrt{\gamma}a^{2}\bar{\Lambda} \right]^{2} \times
\nonumber
\\
&& \quad \qquad \qquad (1 - \gamma a^{2}\bar{\Lambda}^{2}),
 \label{eq:potential2}
\end{eqnarray}
where the $\gamma$ function is obtained from~(\ref{gamma}). The geodetic brane limit is approached when 
$\beta \to 0$, which was deeply studied in~\cite{Davidson:1999fb}. Also, the 
Einstein limit is approached as $\Omega \to 0$ and $\beta \to 0$, which is equivalent to 
$\gamma \to 1$ and $\beta \to 0$.

\begin{figure}[htbp!]
\includegraphics[angle=0,width=6cm,height=6cm]{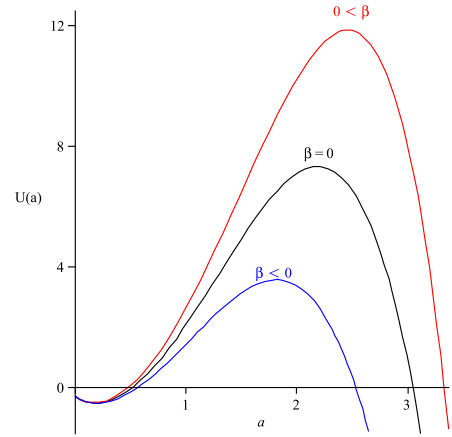}
\caption{Behavior of the WDW type potential for positive, zero and negative values of 
${\beta}$.}
\label{fig:1}
\end{figure}

The parameter $\beta$ which marks the presence of MGBG is still arbitrary at this 
stage. For $\gamma$ real, this potential is well defined for all values of $a$ 
and it exhibits a global maximum in the intermediate region. In fact, this potential has 
a barrier provided  $( \Omega \bar{\Lambda}) ^2 - \left( \frac{2}{3\sqrt{3}} \right)^2
\leq \left( \frac{\bar{\beta}}{\bar{\Lambda}} \right) \left[ 4 (\Omega \bar{\Lambda}) 
 \left( \frac{\bar{\beta}}{\bar{\Lambda}} \right)^2 + \frac{1}{3}  
\left( \frac{\bar{\beta}}{\bar{\Lambda}} \right) + 2 (\Omega \bar{\Lambda}) 
\right]$
where the barrier is stretched between $a_l < a < a_r$ with $a_{l,r}$ being the turning points 
which are the roots of $\bar{\Lambda}^2 a^3 - 3\bar{\beta} a^2 - a + \Omega =0$. 
For the interesting case $\Omega \bar{\Lambda} << 1$ we have that
\begin{eqnarray}
a_l &\simeq& \Omega,
\\
a_r &\simeq& \frac{1}{\bar{\Lambda}} \left[ \left( \frac{3\bar{\beta}}{2\bar{\Lambda}}\right)
+ \sqrt{ \left( \frac{3\bar{\beta}}{2\bar{\Lambda}} \right)^2 + 1 } \right] 
\nonumber
\\
&-& \frac{(\Omega/2)}{ 1 + \left( \frac{3\bar{\beta}}{2\bar{\Lambda}}\right) 
\left[ \left( \frac{3\bar{\beta}}{2 \bar{\Lambda}}\right)
+ \sqrt{ \left( \frac{3\bar{\beta}}{2 \bar{\Lambda}} \right)^2 + 1 } \right] }.
\end{eqnarray} 
In Figure (\ref{fig:1}) we have depicted this potential function. 
This clearly displays that the negative values of the parameter $\beta$ facilitate the creation 
of an expanding universe as the hill of the potential barrier and the turning points are smaller 
in comparison with those obtained by considering the corresponding positive values of $\beta$.
This is in fully agreement with the results obtained at classical level 
reported in~\cite{Cordero:2011zv}, where the self-accelerated expansion of this type of universe 
is owing to $\beta < 0$. There, this parameter plays the role of the crossover scale 
$r_{c}$ in the self-accelerated branch of the DGP model. In addition, at short scale factors
the $a \to 0$ limit implies $\gamma \to \infty$. This gives rise to assume that at
early times the $\gamma$ function can be approximated as $\gamma \simeq  \Omega^{2/3}/ 
(a^{8/3}\bar{\Lambda}^2)$ so that the potential becomes
\begin{equation}
U(a\leq \Omega) \simeq - \Omega^2 - 3\Omega^{4/3} a^{2/3} + 4a^2,
\label{potential3}
\end{equation}
which proves the presence of an embryonic epoch. Note that this expression is insensitive to the
value of $\beta$ and it is similar to the GBG case~\cite{Davidson:1999fb}. This is related to the 
order of approximation that we have used. On the other hand, at long distances the potential
becomes
\begin{eqnarray}
U(a \gg \Omega) &\simeq& 4a^2 \left\lbrace 1 - a^2 \bar{\Lambda}^2 \gamma_0 - \Omega \bar{\Lambda}
\sqrt{\gamma_0} \right. 
\nonumber
\\
&+& \left. \left( \frac{3 \bar{\beta}}{2 \bar{\Lambda}}\right) 
\frac{\Omega}{a^2 \bar{\Lambda} \sqrt{\gamma_0} \left[ \sqrt{\gamma_0}
+ \left( \frac{3 \bar{\beta}}{2 \bar{\Lambda}}\right) \right] }\right\rbrace,
\label{potential4}
\end{eqnarray}
where we have introduced
\begin{equation}
\gamma_0 := 1 + 2 \left( \frac{3\bar{\beta}}{2\bar{\Lambda}} \right)
\left[  \left( \frac{3\bar{\beta}}{2\bar{\Lambda}} \right) - \sqrt{ 
\left( \frac{3\bar{\beta}}{2\bar{\Lambda}} \right)^2
+ 1 } \right].
\label{gamma-0}
\end{equation}
In fact, $\gamma_0$ is the solution to the Eq.~(\ref{gamma}) when $\Omega$ vanishes.
Clearly, for $\Omega \to 0$ and $\beta \to 0$ the potential~(\ref{potential4}) 
approaches to the usual GR quantum potential~\cite{Davidson:1999fb}.
It is immediately to note that we can rewrite the potential~(\ref{potential4})
in terms of an effective cosmological constant as
\begin{equation}
U(a \gg \Omega) \simeq 4a^2 \left( 1 - \Lambda_{\mbox{\tiny eff}} \Omega
- \Lambda_{\mbox{\tiny eff}} ^2 a^2 \right) + U_{\left( \Omega, \Lambda, \beta 
\right)},
\label{potential-6}
\end{equation}
where
\begin{equation}
\Lambda_{\mbox{\tiny eff}} (\Lambda, \beta):= \sqrt{\gamma_0} \bar{\Lambda}, 
\label{eq:Lambda-1}
\end{equation}
and
\begin{equation}
U_{\left( \Omega, \Lambda, \beta 
\right)} = \left(\frac{3\bar{\beta}}{2{\Lambda_{\mbox{\tiny eff}}}}\right)
\frac{ 4 \left( \frac{\Omega}{\Lambda_{\mbox{\tiny eff}}}\right) }{\left[1 + 
\left(\frac{3\bar{\beta}}{
2{\Lambda_{\mbox{\tiny eff}}}}\right) \right]}.
\end{equation}
In particular, we have that $\Lambda_{\mbox{\tiny eff}} (0,\beta) = 0$
and $\Lambda_{\mbox{\tiny eff}} (\Lambda,0) = \bar{\Lambda}$.

\begin{figure}
\centering
\includegraphics[angle=0,width=6.0cm,height=6.0cm]{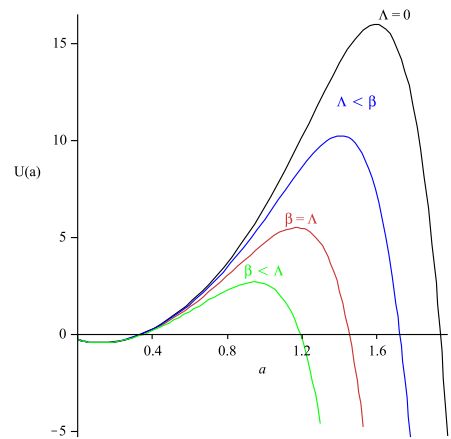}
\caption{Comparison among the WDW type potential for vanishing cosmological constant
with the nonvanishing cases.}
\label{fig2}
\end{figure}

In a like manner, for a vanishing cosmological constant, by a similar development 
we find a potential given by
\begin{equation}
U(a) = a^{2}\left( \gamma a^{2} + 2 + 3\bar{\beta}a^{2}\sqrt{\gamma}\right)^{2}(1-\gamma a^{2}),
\label{pot-2}
\end{equation}
where now the $\gamma$ function satisfies the algebraic equation $\gamma \left( 
\gamma + 3 \bar{\beta}\sqrt{\gamma}\right)^2 = \Omega^2/a^8$. This potential is 
depicted in Figure (\ref{fig2}) for positive values of $\beta$ where, in addition, 
it is compared with those cases where the cosmological constant is non-zero.

\begin{widetext}
\begin{figure*}
\centering
\includegraphics[angle=0,width=6.0cm,height=6.0cm]{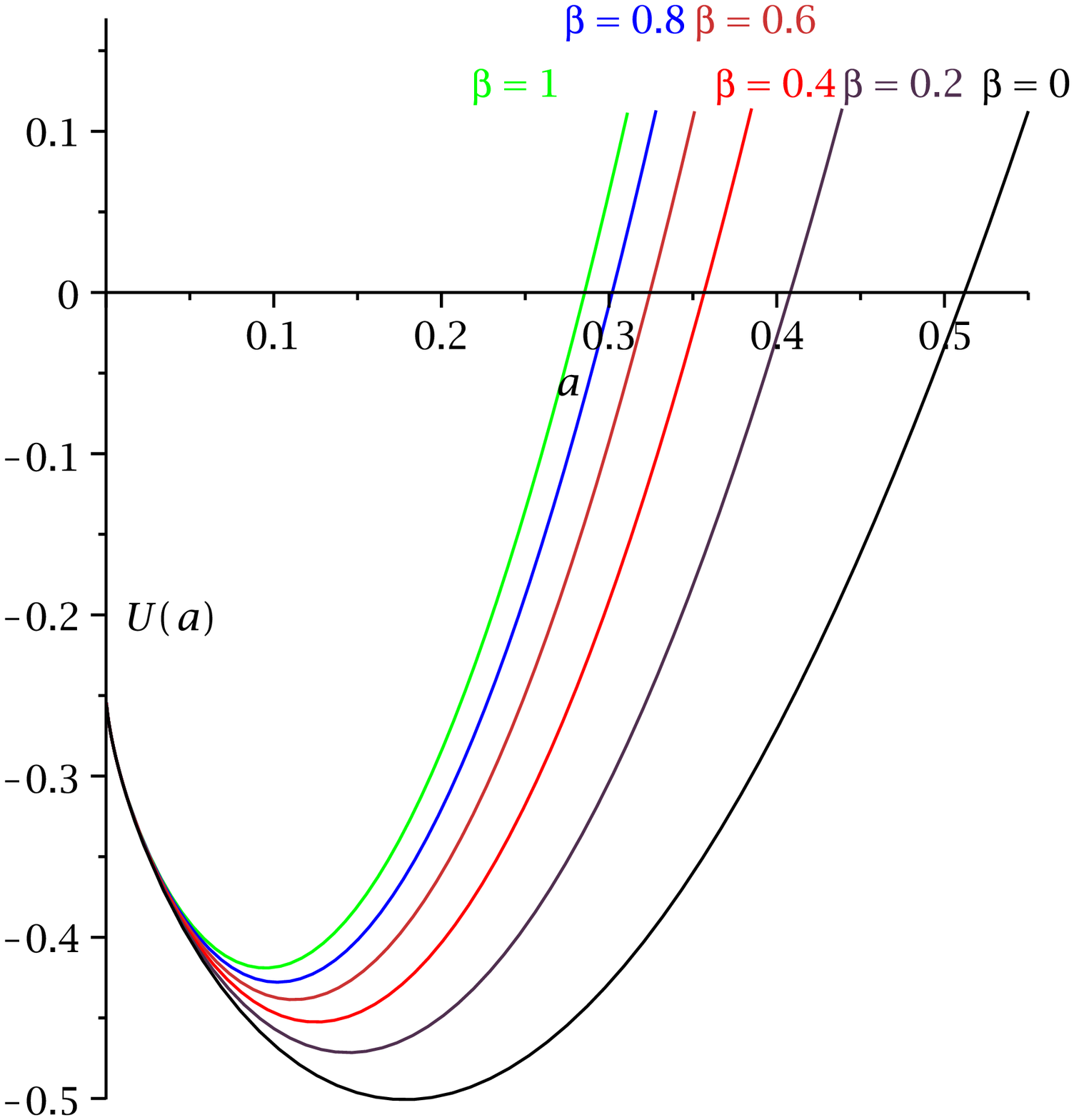}
\includegraphics[angle=0,width=6.0cm,height=6.0cm]{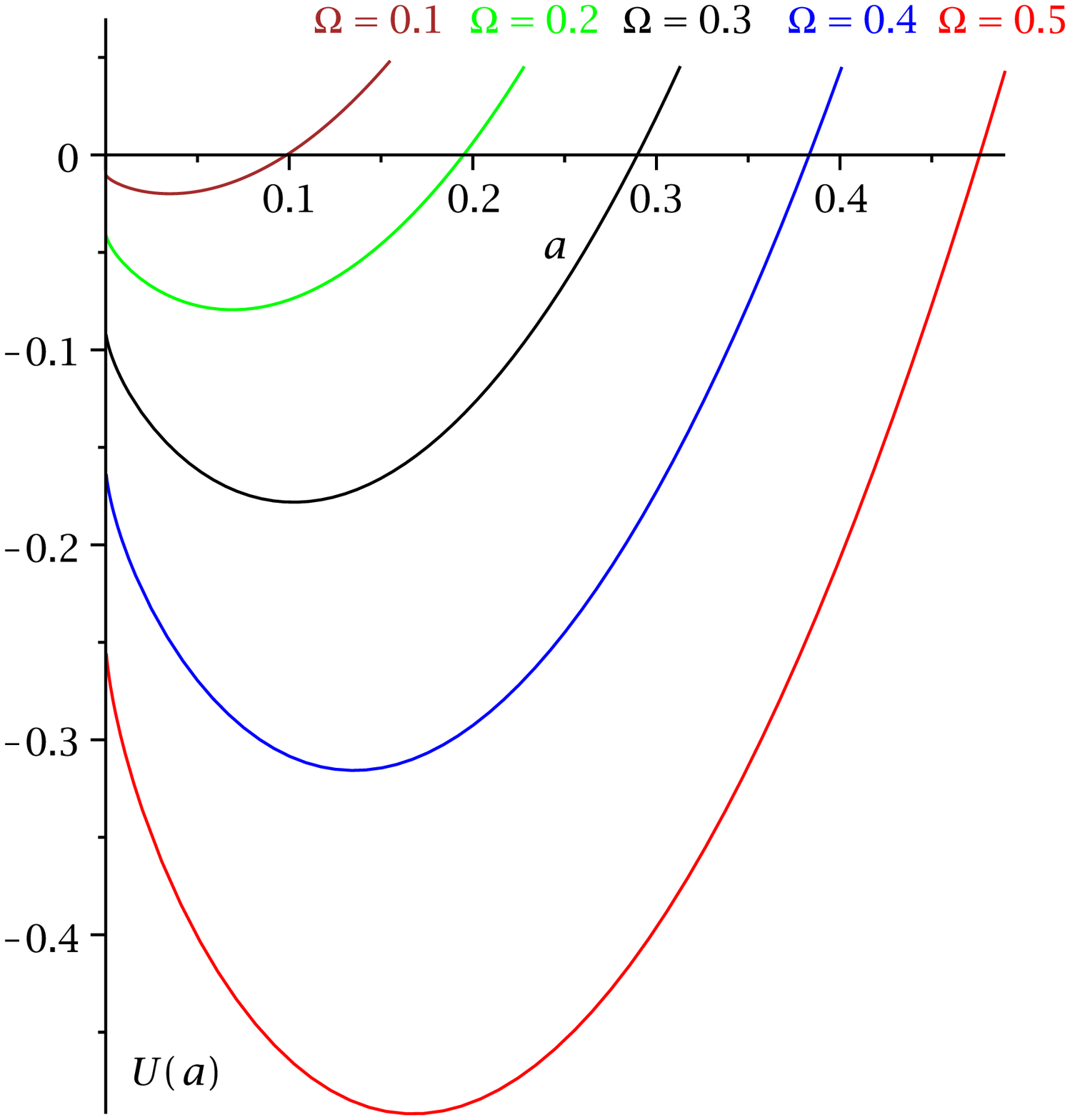}
\caption{\label{fig3}{\it Embryonic epoch}: this region is considered as the deviation 
of the brane-like universe gravity from Einstein gravity~\cite{Davidson:1999fb}. 
The black line is the one obtained from the RT model. Increasing 
values of the parameter ${\bar{\beta}}$ are also considered. For the plot on the 
left side, from yellow to brown 
lines the values correspond to ${\bar{\beta}} = 0.2, 0.4, 0.6, 0.8, 1$, respectively. 
For the plot on the right side, we consider increasing values of the parameter $\Omega$ 
(again, from yellow to brown lines) but maintaining fixed the parameter $\bar{\beta}$.}
\end{figure*}
\end{widetext}

It is expected that the potential function (\ref{pot-2}) may arise from a quantum version
of the model for an accelerated universe without cosmological constant reported in~\cite{Deffayet:2001pu}.
For this brane-like universe we see that the creation from {\it nothing} to a region of unbounded 
expansion is possible and it is privileged whenever we consider a cosmological constant on 
the brane. Moreover, for small values of the parameter ${\beta}$ and $\Lambda = 0$ the 
potential barrier grows rapidly making harder the analysis of the tunneling effects. We 
further observe, as long as $\Omega \neq 0$, for the range of small values for $a$ we 
still have an embryonic epoch because in such regions the Universe can exist classically. 
In fact, the embryonic epoch takes place whenever the brane energy $\Omega \neq 0$ which 
is the main element of the unified brane gravity~\cite{DG}. In this regard, Fig.~(\ref{fig3}) 
shows that the embryonic region is bigger for small values of ${\beta}$ and large values of $\Omega$.

%
\section{Nucleation Rate}
\label{sec6}
%

In the quantum cosmology framework the whole universe is described by a wavefunction. 
The question of the right boundary conditions for the wavefunction is hard to answer 
because, unlike ordinary quantum mechanics where boundary conditions for the wavefunction 
are fixed by the physical set-up external to the system, in 4D quantum cosmology there 
is nothing external giving as a consequence that this question does not have a clear 
resolution~\cite{debate}.
In our case, the existing embedding spacetime makes the main difference. This is so because 
the presence of the bulk space gives, without ambiguity, the following interpretation: the 
Hartle-Hawking and Linde boundary conditions include parts that correspond to expanding and 
contracting universe whereas the tunneling boundary condition only includes an expanding 
component for the Universe (see the discussion, for example, in \cite{stealth}).

We opt to think that this brane-like universe was a small nearly spherical brane nucleating in a 
Minkowski background spacetime and we choose the tunneling boundary condition as the 
right boundary condition because it corresponds to the idea that the tunneling mechanism  
was the process involved in the nucleation of this universe.

For our case, by a WKB approximation it is possible to calculate the nucleation probability 
considering the tunneling boundary condition driven by the involved potential~(\ref{eq:potential2}) 
as follows~\cite{Garriga:1993fh,Vilenkin:1984wp,Vilenkinap}
\begin{equation}
\mathcal{P} \sim \exp\left(- 2\int^{a_{r}}_{a_{l}}|\sqrt{U(a)}|da\right),
\label{eq:nucle}
\end{equation}
where, ${a_{l}}$ and ${a_{r}}$ are the turning points of the potential. 
Clearly, this expression is hard to work out.

A special case focused on the very early Universe is contained in the case $\Omega = 0$. 
From Eq.~(\ref{gamma}) the $\gamma$ function reduces to $\gamma_0$. 
Under this condition the effective quantum potential~(\ref{eq:potential2}) takes the form
\begin{equation}
U(a) = 4a^2 \left[ 1 - a^2 \left( \sqrt{\gamma_0}\bar{\Lambda}\right)^2 \right]. 
\label{potential5}
\end{equation}
The turning points becomes
\begin{eqnarray}
a_l &=& 0,
\\
a_r &\simeq& \frac{1}{\bar{\Lambda}} \left[  \left( \frac{3\bar{\beta}}{2\bar{\Lambda}} 
\right) + \sqrt{ \left( \frac{3\bar{\beta}}{2\bar{\Lambda}} \right)^2 + 1 } \right].
\end{eqnarray}
Thus, from~(\ref{eq:nucle}) we may estimate the tunneling probability
\begin{equation}
\mathcal{P} \sim \mbox{e}^{-\frac{4}{3\left( \sqrt{\gamma_0}\bar{\Lambda}\right)^{2}}}.
\label{probability}
\end{equation}
Notice that both the potential (\ref{potential5}) and the nucleation probability~(\ref{probability}) 
resemble the standard GR case with an effective cosmological constant defined in~(\ref{eq:Lambda-1}),
\begin{equation}
{\Lambda}_{\mbox{\tiny{eff}}} = \sqrt{\gamma_0}\bar{\Lambda} =  \sqrt{\bar{\Lambda}^{2} + 
\left( \frac{3\bar{\beta}}{2}\right)^2} - \frac{3\bar{\beta}}{2}.
\end{equation}
From expression (\ref{probability}), we also infer that it is more probable to create universes  
of this type with a value of $\bar{\Lambda}$ greater than the usual situation of GR but with the
main difference that this effect can be increased by considering negative
values of the parameter $\beta$. The opposite situation occurs whenever ${\beta}>0$. 

Associated with this case, when ${\Lambda}=0$ and $\Omega = 0$, and by using the 
potential~(\ref{pot-2}) we have that
\begin{equation}
U(a) = 4 a^2 \left( 1 - 9 \bar{\beta}^2 a ^2 \right). 
\end{equation}
From this expression we may also identify another effective cosmological constant 
given by $\bar{\Lambda}_{{\mbox{\tiny eff}}} := -3\bar{\beta}$. These results together 
suggest that in general, at quantum level, the parameter ${\beta}$ still continues 
to modify the cosmological constant, as it was elucidated at a classical level 
in~\cite{Cordero:2011zv}.

%
\section{Concluding remarks}
\label{sec7}
%

In this paper we have canonically quantized the modified Regge-Teitelboim brane 
model within a minisuperspace framework. The associated brane-like universe also
constitutes a controlled deviation from Einstein limit provided the bulk energy
$\Omega$ and the $\beta$ parameter vanish. By means of an Ostrogradski Hamiltonian procedure 
besides the introduction of a suitable canonical transformation followed of a gauge
fixing procedure, we have succeeded in finding constraints quadratic in the momenta. 
The canonical quantization scheme is possible once the Dirac brackets enter the
game. The resulting WDW type equation allows to identify a quantum
potential. We calculated then the nucleation probability using the WKB approximation 
for the simple case $\Omega = 0$. For this case, the higher probability for the 
nucleation of the brane universe is obtained by considering the highest value of the resulting 
effective cosmological constant that is constructed with the cosmological constant 
$\Lambda$ and the $\beta$ parameter. For the case of positive $\beta$ there is a 
relation between $\beta$ and $\Lambda$ in order to get the maximum probability. For 
$\beta < 0$ the maximum probability is achieved for the largest absolute values of $\beta$ 
and $\Lambda$. For a non-zero brane cosmological constant, the 
parameter $\beta$ which plays akin role to the crossover scale $r_{c}$ in 
the self-accelerated branch of the DGP model, is similar to a cosmological constant 
at a quantum level. In some manner, this is consistent with the results reported 
in~\cite{Cordero:2011zv}, where the self-accelerated expansion on this type of universes 
is due to a negative value of the parameter $\beta$. Based on the previous 
results, it is worth to mention that exist the possibility that the nucleation 
probability for the DGP model follows a similar pattern.

Another way to extend our work resides in the direction of trying to extract some 
physical observable consequences from the nucleation rate. For example, we can 
use an inflaton field on the brane, within a particular inflationary model, with 
the purpose to calculate some of most probable observable cosmological parameters and to 
investigate its corresponding physical implications.This will be the subject of 
future work.

We believe also that this model will serve as precursor to obtain more enriched 
geometrical theories. In this sense, Lovelock brane models~\cite{Cruz:2012bw} 
are guesses at alternative physical theories that might underlie the cosmic 
acceleration that deserve a more detailed exploration. We will report elsewhere 
the cosmological implications of consider, for example,  a cubic correction term 
in the extrinsic curvature to the RT model through the so-called Gibbons-Hawking-York-Myers 
term, ${\cal K}_{GB}$~\cite{myers,davis}.

\acknowledgments
The authors are grateful to Alexander Vilenkin for helpful discussions. AM 
acknowledges support from PROMEP UASLP-PTC-402. ER acknowledges partial support 
from grant PRO\-MEP, CA-UV: Algebra, Geometr\'\i a y Gravitaci\'on. MC was supported 
by PUCV through Proyecto DI Postdoctorado 2014. Also, ER and MC acknowledge partial 
support from the grant CONACyT CB-2012-01-177519. This work was partially supported 
by SNI (M\'exico). RC also acknowledges support from EDI, COFAA-IPN, SIP-20131541 
and SIP-20144150.


\begin{thebibliography}{}

\bibitem{Cordero:2011zv}
R. Cordero, M. Cruz, A. Molgado and E. Rojas, Classical and Quantum Gravity {\bf 29}, 
175010 (2012).

\bibitem{Regge}
T.~Regge and C.~Teitelboim,
{\it Proceedings of the First Marcel Grossman Meeting}, (North-Holland, Amsterdam, 1977).

\bibitem{pavsic} 
M. Pav\v{s}i\v{c}, Phys. Lett. A {\bf 107} 66 (1985).

\bibitem{davidson0} 
A. Davidson and D. Karasik, {\it Mod. Phys. Lett. A} {\bf 13}, 2187 (1998).

\bibitem{davidson1} 
D. Karasik and A Davidson, {\it Phys. Rev. D} {\bf 67}, 064012 (2003).

\bibitem{paston2} 
S. A. Paston, A. A. Sheykin, ``The approach to gravity as a theory of embedded surface'', 
\textit{arXiv}: 1402.1121 [gr-qc]

\bibitem{DGP} 
G.~R.~Dvali, G.~Gabadadze and M.~Porrati, Phys. Lett. B {\bfseries 484}, 112 (2000);
{\bfseries 485}, 208 (2000).

\bibitem{DGP1} 
G.~R.~Dvali, G.~Gabadadze and M.~Porrati, ``4D gravity on a brane in 5D Minkowski space'', Phys. Lett. B {\bfseries 485}, 208 (2000);
{\bfseries 485}, 208 (2000).

\bibitem{chen} 
B-Y, Chen, J. London Math. Soc. {\bf 6} 321 (1973)

\bibitem{svetina} 
S. Svetina and B, \v{Z}ek\v{s}, Eur. Biophys. J {\bf 17} 101 (1989)

\bibitem{ot} 
M. \"{O}nder and R. M. Tucker, J. Phys. A: Mathematical and General {\bf 21} 
3423 (1988); M. \"{O}nder and R. W. Tucker, Phys. Lett. B {\bf 202} 501 (1988)

\bibitem{electron} 
R. Cordero, A. Molgado and E. Rojas, Class. Quant. Grav. {\bf 28} 065010 (2011)

\bibitem{davidson3} 
A. Davidson and S. Rubin, Class. Quant. Grav. {\bf 28} 125005 (2011)

\bibitem{trodden1} 
G. L. Goon, K. Hinterbichler and M. Trodden, Phys. Rev. Lett. {\bf 106} 231102 (2011) 

\bibitem{trodden3} 
G. L. Goon, K. Hinterbichler and M. Trodden, M. J. Cosmology Astrop. Phys. {\bf 07} 
017 (2011)

\bibitem{Davidson:1999fb} 
A. Davidson, D. Karasik and Y. Lederer, Class. Quant. Grav. {\bf 16} 1349 (1999)

\bibitem{friedman:1965} 
A. Friedman, A. Rev. Mod. Phys. {\bf 37} 201 (1965)

\bibitem{rosen:1965} 
J. Rosen, Rev. Mod. Phys. {\bf 37} 204 (1965)

\bibitem{Cruz:2012bw}
M.~Cruz and E.~Rojas, Classical and Quantum Gravity {\bfseries 30}, 115012 (2013). 

\bibitem{hambranes} 
R. Capovilla, J. Guven and E. Rojas, Class. Quant. Grav. {\bf 21} 5563 (2004)

\bibitem{Nicolis:2008in} 
A. Nicolis, R. Rattazzi and E. Trincherini, Phys. Rev. D {\bf 79} 064036 (2009)

\bibitem{deRham:2010eu}
C. de Rham and A. J. Tolley, J. Cosmology Astrop. Phys. {\bf 05} 015 (2011)

\bibitem{Goon:2011xf} 
G. Goon, K. Hinterbichler and M. Trodden, J. Cosmology Astrop. Phys. {\bf 12} 004 
(2011)

\bibitem{Burrage:2011bt}
C. Burrage, C. de Rham and L. Heisenberg, J. Cosmology Astrop. Phys. {\bf 05} 025 (2011)

\bibitem{Deffayet:2009mn}
C. Deffayet, S. Deser and G. Esposito-Farese, Phys. Rev. D {\bf 80} 064015 (2009)

\bibitem{Deffayet:2009-2}
C. Deffayet, G. Esposito-Farese and A. Vikman, Phys. Rev. D {\bf 79} 084003 (2009)

\bibitem{Deffayet:2010}
C. Deffayet, S. Deser and G. Esposito-Farese, Phys. Rev. D {\bf 82} 061501(R) (2010)

\bibitem{Fairlie:1992} 
D. B. Fairlie, J. Govaerts and A. Morozov, Nucl. Phys. B {\bf 373} 214 (1992)

\bibitem{TRODDEN2} 
M. Trodden and K. Hinterbichler, Class. Quant. Grav. {\bf 28} 204003 (2011)

\bibitem{Cordero:2009mj} 
R. Cordero, A. Molgado and E. Rojas, Phys. Rev. D {\bf 79} 024024 (2009)

\bibitem{Paul2013} 
B. Paul, Phys. Rev. D {\bf 87} 045003 (2013)

\bibitem{Paul2013-2} 
R. Banerjee, P. Mukherjee and B. Paul, Phys. Rev. D {\bf 89} 043508 (2014)

\bibitem{paston1} 
S. A. Paston and A. N. Semenova, Int. J. Theor. Phys. {\bf 49} 2648 (2010)

\bibitem{davidson} 
A. Davidson, Class. Quant. Grav. {\bf 16} 653 (1999)

\bibitem{Dirac} 
P. A. M. Dirac, Lectures on Quantum Mechanics. Dover publications, Mineola, New York, 
(2001)

\bibitem{Henneaux} 
M. Henneaux and C. Teitelboim, Quantization of gauge systems. Princeton University 
Press, Princeton, New Jersey (1992)

\bibitem{Nesterenko1989} 
V. V. Nesterenko, J. Phys. A: Mathematical and General {\bf 22} 1673 (1989)

\bibitem{Ostro:1850} 
M. Ostrogradski, Mem. Ac. St. Petersbourg {\bf VI 4} 385 (1850)

\bibitem{tolley} 
T. Chen, M. Fasiello, E. A. Lim and A. Tolley, J. Cosm. Astrop. Phys. {\bf 02} 
042 (2013)

\bibitem{Ho2003} 
P. M. Ho, Phys. Lett. B {\bf 558} 238 (2003)

\bibitem{Deffayet:2001pu} 
C. Deffayet, G. R. Dvali and G. Gabadadze, Phys. Rev. D {\bf 65} 044023 (2002)

\bibitem{DG} 
A. Davidson and I. Gurwich, Phys. Rev. D {\bf 74} 044023 (2006)

\bibitem{debate} 
A. Vilenkin, Phys. Rev. D {\bf 58} 067301 (1998)

\bibitem{stealth} 
R. Cordero and A. Vilenkin, Phys. Rev. D {\bf 65} 083519 (2002)

\bibitem{Garriga:1993fh} 
J. Garriga, Phys. Rev. D {\bf 49} 6327 (1994)

\bibitem{Vilenkin:1984wp} 
A. Vilenkin, Phys. Rev. D {\bf 30} 509 (1984)

\bibitem{Vilenkinap} 
A. Vilenkin, Phys. Rev. D {\bf 50} 2581 (1994)

\bibitem{myers} 
R. C. Myers, Phys. Rev. D {\bf 36} 392 (1987)

\bibitem{davis} 
S. C. Davis, Phys. Rev. D {\bf 67} 024030 (2003)

\end{thebibliography}
\end{document}